\documentclass[11pt]{article}

\usepackage[margin=1in]{geometry}
\usepackage{microtype}
\usepackage{fancyhdr}
\usepackage[T1]{fontenc}
\usepackage{lmodern}

\usepackage{graphicx}
\usepackage{float}
\usepackage{subcaption}
\usepackage{tabularx}
\usepackage{booktabs}
\usepackage{tikz}
\usetikzlibrary{arrows.meta,positioning,shapes.geometric}

\newcolumntype{Y}{>{\raggedright\arraybackslash}X}

\usepackage[most]{tcolorbox}
\newtcolorbox{claimbox}[1][]{
  colback=blue!5!white,
  colframe=blue!50!black,
  boxrule=0.8pt,
  arc=2pt,
  left=8pt,
  right=8pt,
  top=8pt,
  bottom=8pt,
  fonttitle=\bfseries,
  title=Core Claim,
  #1
}

\usepackage{amsmath,amssymb}

\usepackage[hidelinks]{hyperref}

\usepackage[numbers,sort&compress]{natbib}

\usepackage{enumitem}
\setlist[itemize]{leftmargin=*,topsep=3pt,itemsep=2pt,parsep=0pt}
\setlist[enumerate]{leftmargin=*,topsep=3pt,itemsep=2pt,parsep=0pt}

\title{\bf When Scientific Cognition Is No Longer Scarce
}
\author{
Nathan DeBardeleben \\
Los Alamos National Laboratory \\
\small
\texttt{ndebard@lanl.gov}
}
\date{}

\begin{document}
\maketitle

\begin{abstract}
AI could change which parts of science impede progress.  Consider a world in
which machine systems are better, faster, and cheaper than people at most
scientific work \emph{that can be done through a computer}.  Our question is
what
would limit science in that world.  Literature synthesis, hypothesis
generation, software development, simulation, and analysis could become
abundant, while experiments, observations, well-supported conclusions, and
accountable institutional authority remain scarce.  Science would then be
constrained by a different set of resources.  In this paper, we call this
change the
\emph{scarcity inversion} and consider four parts of it: selection, physical
access, validation, and organizational choice.  This change is arriving
first in mathematics and coding/software/algorithm design, where the whole
scientific loop can run inside computation.  For national laboratories, the
change could be striking.  Their distinctive role is to turn abundant machine
reasoning into trustworthy results by combining controlled experiments,
protected data, expert judgment, and accountable authority.  The practical
question is how
facilities, verification, provenance, resource allocation, and scientific
governance should change when reasoning is plentiful and trustworthy evidence
is scarce.
\end{abstract}

\section{Introduction: What If Scientific Cognition Becomes Abundant?}
Let's start with a premise
that AI becomes better, faster, and cheaper than people at
nearly every scientific task that can be done with access to a computer.  Set
aside whether this happens, or when.  Take the premise seriously long enough to
ask one question: what limits scientific progress when scientific
\emph{cognition} is no longer scarce?

We examine how science changes when the supply of reasoning, coding, and
analysis expands enough that these activities cease to limit scientific
progress.

To make the question concrete, we will use a simple sketch of the scientific
workflow:

\begin{itemize}
  \item synthesizing literature and existing evidence
  \item generating hypotheses and candidate explanations
  \item designing experiments, observations, and simulations
  \item developing software, models, and experimental procedures
  \item executing simulations and interacting with the physical world
  \item analyzing results and comparing competing explanations
  \item validating conclusions, characterizing uncertainty, and determining
    which claims the evidence supports
  \item communicating findings so they can be scrutinized, reproduced, and
    incorporated into future work.
\end{itemize}

In practice, scientific work loops back on itself.  A failed experiment may
expose a flaw in the design.  An unexpected result may instead require a new
hypothesis or another look at the literature.  The workflow sketch lets us ask
where AI speeds up the loop and where scarcity remains.

Mathematics and parts of theoretical computer science can keep this entire
workflow inside computation.  They offer an early view of what happens when
the search, execution, and decisive checks can all proceed at machine speed.
At Los Alamos National Laboratory (LANL), this is already a practical
question.  Scientific AI work at the Laboratory gives us a direct view of both
the opportunities and the constraints.

\textbf{Our claim is that abundant cognition changes what is scarce in science.}
Selection, physical evidence, validation, trust, and accountable institutional
authority become the harder constraints.

The rest of this paper is organized as follows.
Section~\ref{sec:scarce-expertise} describes how science is organized around
scarce expertise, and
Section~\ref{sec:cognitive-bottleneck} considers what changes when reasoning is
no longer the limiting resource.  Section~\ref{sec:compute-only} examines
fields where the whole scientific loop fits inside computation, and
Section~\ref{sec:scarcity-inversion} introduces the concept of a scarcity
inversion.
Sections~\ref{sec:selection} through
\ref{sec:institutions} examine the resulting bottlenecks in selection, physical
access, validation, and institutional choice.  Section~\ref{sec:labs}
considers the changing role of national laboratories, and
Section~\ref{sec:design} discusses how scientific systems might be designed
around the resources that remain scarce.  Finally, we conclude in
Section~\ref{sec:conclusion}.

\section{Science Organized Around Scarce Expertise}
\label{sec:scarce-expertise}

Today, expert time limits every part of scientific work.  A day spent reading
or inspecting results cannot also be spent testing another hypothesis or
writing code.  Institutions therefore ration cognitive work along with
facilities and funding.

Much of scientific apprenticeship grows out of delegated work.  A senior
scientist may ask a junior colleague to run a simulation campaign and return
with an interpretation.  Doing that work develops the methods and judgment
needed for independent research.

Apprenticeship also teaches scientists what deserves a second look.  Repeated
contact with instruments, failed runs, and imperfect data builds the background
against which an anomalous result registers as surprising.
If routine work moves to agents, institutions will need deliberate ways for
junior scientists to develop that judgment.

Experimental work also contains judgment that cannot be separated cleanly
from execution.  Scientists decide whether an instrument is behaving normally,
whether a specimen is representative, whether an anomaly deserves attention,
and whether a measurement bears on the question being asked.  Automation can
change who or what operates the apparatus, but it does not remove the need for
people who understand the experiment and its failure modes.

Matt Beane develops this concern in \emph{The Skill Code}, drawing on studies of
surgical suites, warehouses, and other workplaces \citep{beane2024skillcode}.
Richard Mitchell draws a practical lesson from aviation and nuclear control:
organizations can put selected ``manual gates'' into automated workflows so
people continue to practice skills that the organization cannot afford to lose
\citep{mitchell2026aiefficiency}.
A broader labor-market analysis projects that generative AI may automate
foundational work in roles such as legal associate, narrowing the entry points
through which novices develop expertise \citep{sigelman2025expertise}.  Recent
payroll data show a widening employment gap for workers ages 22--25 in
AI-exposed occupations, driven mainly by lower hiring.  The gap is largest in
occupations where reported AI use primarily automates work
\citep{brynjolfsson2026canaries}.  The authors treat this as descriptive
evidence and do not claim a causal estimate.  What this does to the next
generation of skilled scientists deserves its own essay, and we leave it aside
here.

Scientific institutions (universities, national laboratories,
science divisions of companies)
are organized around a limited supply of expert time.
Laboratories concentrate expertise, divide work among specialized teams, and
plan projects around what those teams can do.  If expert time stops being the
main constraint, laboratories, teams, and career paths will have to adjust.

\section{Removing the Cognitive Bottleneck}
\label{sec:cognitive-bottleneck}

If scientific reasoning becomes abundant, several historical
constraints seem to weaken at the same time.
Literature synthesis and related-work analysis may become continuous
activities, making hypothesis generation much more fluid as well as cheaper
than humans.
Every model
(LLM, surrogate, emulator) call,
retrieval step, simulation, and round of agent debate still costs money
(time, energy, compute cycles).
Scientific agents therefore face a selection problem
before any candidate reaches a laboratory: where should they spend the next
LLM token, tool call, or simulation?  The choice should reflect plausibility,
cost, and how much a result would reduce uncertainty.  A simulation that
resolves a large branch of later work may deserve priority even when its
immediate result has little value on its own.

Selection begins within hypothesis generation itself.  Pal and colleagues
studied a graph-based system over as many as 2,000 iterations.  After a few
hundred iterations, new concepts were no longer taking the graph farther from
its starting point, although the number of distinct concepts kept growing.
The system continued to find new long-range connections among those concepts
\citep{pal2026graphnative}.  Further compute was now exploring connections
among ideas already in view.  That shift creates a stopping problem.  At some
point, a fresh starting point or an empirical test may be worth more than
another iteration.

Scientific software could become so cheap to produce and modify that we may
come to think of it as disposable, upending the systems institutions use to
manage software as intellectual property: release review, licensing, controlled
distribution, and technology transfer.  In many ways, we are already
seeing this with a proliferation of software agentically created under the
direction of scientists.
Design of experiments and simulation campaigns could proliferate far
faster than researchers can execute them, let alone at the rate humans can
digest the results.

Internal measurements from OpenAI give an early example in AI research.  By
mid-August 2026, its research organization was using 3.1 eight-hour agent
workdays for every human workday.  OpenAI also reports faster code contributions
and a record number of experiments per active experimenter.  The measurements
are preliminary, increased compute may partly explain the experiment trend, and
throughput does not necessarily translate directly into scientific progress
\citep{openai2026researchacceleration}.

Sakana AI's AI Scientist automates a software-only research loop from an
initial idea to a draft paper and simulated review.  Along the way it writes
code, runs experiments, and analyzes the results.  It currently focuses on
machine learning experiments, where the experimental loop can be carried out
entirely through software \citep{lu2024aiscientist}.

Google's Co-Scientist applies agent debate and ranking to biomedical
hypotheses.  Researchers evaluated the system on drug repurposing,
novel-target discovery, and
antimicrobial resistance, then tested selected proposals in wet-laboratory
experiments.  Experts remained in the loop to choose what received laboratory
time \citep{gottweis2026coscientist}.

Los Alamos National Laboratory's agentic framework, URSA
\citep{grosskopf2026ursa} (Universal Research and Scientific Agent), takes
similar approaches to automation.  Scientists can easily explore parameter
spaces, run a suite of simulations, build surrogate models from results,
analyze results, generate plots, etc.
URSA has a suite of cognitive features such as a healthy debate/critique
loop as well as a symposium of agents that specialize with different roles and
tools and collaborate to solve complex problems.

Cheap generation can fill the workflow with plausible ideas and designs.  The
difficulty then lies in deciding which candidates deserve further analysis,
simulation, or experiment.

\section{Science That Fits Entirely Inside a Computer}
\label{sec:compute-only}

Much of this paper focuses on the point where scientific work needs an
experiment, an observation, or access to a facility (experimental or
measurement device/instrument).  Some problems have no
such constraint.  In pure mathematics, formal methods, and parts of theoretical
computer science, the problem, candidate solutions, and decisive checks can
all be represented and executed in a computer.  These fields give us an early
view of what happens when the whole loop can run at machine speed.

FunSearch supplied an early example that paired LLM-generated programs with an
automatic evaluator and found new constructions for the cap set problem,
including the largest improvement in 20 years to the asymptotic lower bound.
It also found new heuristics for online bin packing
\citep{romeraparedes2024funsearch}.  AlphaEvolve later used an autonomous loop
of LLM code changes and automatic evaluation across a wider set of problems.
Among its results was a procedure for multiplying two $4 \times 4$
complex-valued matrices using 48 scalar multiplications, the first improvement
over Strassen's method in that setting in 56 years
\citep{novikov2025alphaevolve}.

\begin{figure}[H]
  \centering
  \includegraphics[width=\textwidth]{%
    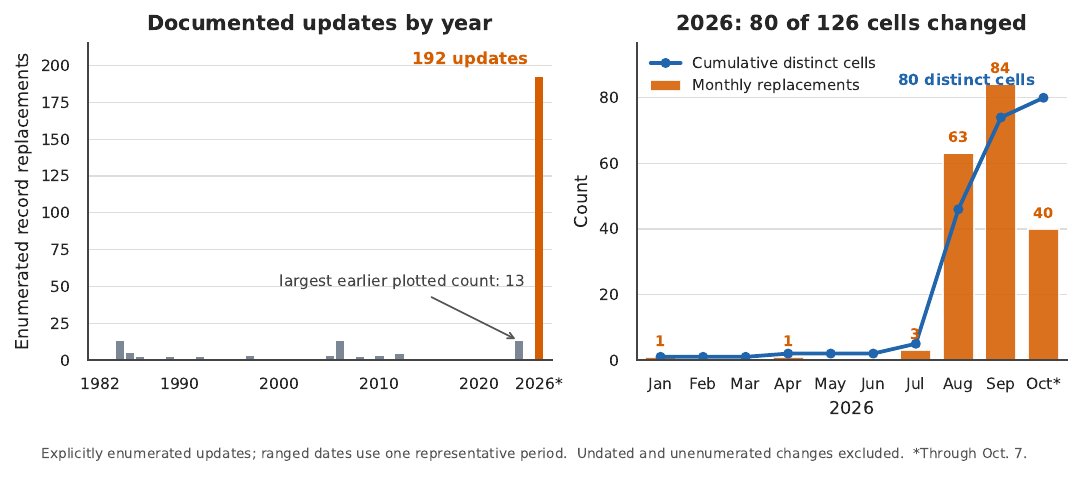}
  \caption{Documented turnover across all 126 degree--diameter cells.  The
    left panel counts explicitly enumerated record replacements by year.  The
    right panel expands 2026 and tracks the cumulative number of distinct cells
    changed\citep{comellas2026degdiameter}.}
  \label{fig:degree-diameter-turnover}
\end{figure}

The degree--diameter problem gives a concrete view of this acceleration.  A
community-maintained table tracks the largest known graph for 126 combinations
of degree and diameter.  Its update log enumerates 192 record replacements in
2026 through October 7, touching 80 of the 126 cells.  This burst coincides
with the arrival of the AI-assisted systems described throughout this section.
Historical reporting is incomplete, so the comparison does not measure every
advance or establish that AI caused the increase.  Even with that limitation,
the breadth and concentration of the turnover show how fast a compute-only
field can move when candidate generation and exact checking both accelerate
\citep{comellas2026degdiameter}.
Figure~\ref{fig:degree-diameter-turnover} shows both the abrupt rise in
documented updates and the breadth of the 2026 turnover.

These systems still consume finite compute.  Their advantage comes from
keeping generation and evaluation in the same medium.  The result of one
candidate can feed directly into the next round of search.  No physical
experiment interrupts the loop.

Recent OpenAI work in mathematics makes the trend harder to dismiss.  GPT-5
contributed to progress on several Erd\H{o}s problems through literature
search, proof generation, and expert correction
\citep{bubeck2025earlyscience}.  Subsequent releases included a disproof of the
Erd\H{o}s unit distance conjecture, ten further results formalized in
Lean\footnote{Lean was provided as a tool for the AI agents to use, effectively
becoming part of the verification loop.}, and a claimed proof of finite-time
blowup in the three-dimensional incompressible Navier--Stokes equations
produced with as many as $10,000$ agents
\citep{openai2026unitdistance,alon2026unitdistance,
openai2026tenadvances,openai2026navierstokes}.

In October 2026, OpenAI released a collection of 722 manuscripts organized
into 372 result families.  The collection was produced by an unreleased
internal model after it was posed approximately 4,000 open problems.  OpenAI
reports that an average result used the equivalent of roughly three hours of
ChatGPT Pro thinking.  Many proofs have Lean formalizations, while
unformalized results may still contain errors
\citep{openai2026mathcatalog}.  These releases make the scarcity inversion
visible.  Candidate mathematical results are being produced at scale, while
formalization, independent review, correction, and assimilation into the
mathematical record become the limiting work.  The search and checking take
place largely inside computation, without a physical facility or a standing
team of specialists for each problem.

This part of the scarcity inversion should concern national laboratories
first.  Laboratories have long employed mathematicians, computer scientists,
and algorithm developers to do this work.  When a compute-only problem relies
on public information, it draws no advantage from a beamline, controlled
material, or unique location.  Models, compute, formal tools, and expert review
can be assembled elsewhere.  This is less true for National Nuclear Security
Administration (NNSA) laboratories.  They hold classified and controlled data,
models, and technical records that cannot be assembled elsewhere.  AI may make
those collections easier to curate, search, and use, but their value still
depends on secure access, provenance, and institutional knowledge.  National
laboratories can also connect abstract results to consequential physical
systems.  Their leverage increasingly comes from these protected information
resources and from their connection to the physical world.

\section{The Scarcity Inversion}
\label{sec:scarcity-inversion}
In this hypothetical future, candidate knowledge can grow much faster than
validated knowledge.  Machine reasoning will still consume energy, hardware,
and time, and difficult problems will continue to resist solution.  The
inversion concerns relative scarcity: another hypothesis, analysis, or software
implementation becomes cheap compared with the experiment needed to test it.

\begin{claimbox}[title=The Scarcity Inversion]
As machine reasoning becomes abundant, the scarce resources in science shift
toward selecting worthwhile questions, obtaining contact with the physical
world, determining which claims the evidence supports, and exercising
accountable institutional authority.
\end{claimbox}

A serious objection follows from the premise itself.  Capable agents should
also make selection and validation cheaper because both involve reasoning.
The remaining limit appears when a task depends on something inference cannot
supply.  An unmade measurement is unavailable to a model.  Authority and
public legitimacy must come from an institution.  Agents can improve the
analysis behind these decisions, but the scarce resource lies outside
computation.

The four scarcities constrain different parts of the same scientific loop:

\begin{itemize}
  \item Selection determines which questions are well-developed and which
    proposals receive further attention.
  \item Physical access determines which proposals can be tested through an
    experiment, measurement, or observation.
  \item Validation determines which claims the resulting evidence supports and
    how confidently scientists and engineers can make them.
  \item Institutional choice determines whether those claims justify action,
    continued investment, or an accepted risk, and identifies who is
    accountable for the decision.
\end{itemize}

Human institutions shape conduct through norms and through accountability.
Accountability makes misuse of authority consequential for the actor's future.
Alignment may shape an agent's conduct, but present agents cannot be held
accountable in that institutional sense.  The practical importance of this
\emph{accountability asymmetry} grows as systems gain authority to change
experiments, infrastructure, or resource allocations
\citep{debardeleben2026accountability}.

Some scientific decisions still need a person or institution that can be held
responsible for the outcome.  Giving an agent operational control does not
transfer that responsibility.

\section{The Selection Bottleneck: What Is Worth Knowing?}
\label{sec:selection}

When a system can generate millions of plausible hypotheses and thousands of
defensible experimental designs, science faces a selection problem.  Which
question deserves scarce resources, and which contact with reality would be
most valuable next?  Information gain, uncertainty reduction, mission
relevance, risk, and cost all enter that choice.

This choice begins inside the computational workflow.  Candidate hypotheses
compete on plausibility, novelty, feasibility, and expected information value.
Those criteria can disagree.  In their experiments, O'Neill, Ghosal, and
colleagues found that fine-tuned models produced hypotheses that an LLM judge
scored as more feasible and less novel than one-shot variants
\citep{oneill2025sparks}.  Their recent work on hypothesis ranking also shows
how difficult automated selection remains.  On this benchmark,
confidence-based scoring outperformed a prompted LLM judge, but performance
remained far from reliable and the benchmark used previously published
hypotheses \citep{rajwal2026goodhypothesis}.  A system that can generate far
more than it can examine needs an explicit policy for allocating its own
reasoning budget.

Selection occurs twice.  The first decision allocates machine effort among
searches, simulations, and analyses.  The second admits a small fraction of the
resulting proposals to facilities, field campaigns, or human review.
Computational search can explore broadly because a mistake is comparatively
cheap.  Facility admission needs a stronger account of expected value because
it commits resources that cannot be copied or recovered as easily.

Ranking candidates one at a time can hide a weakness at the portfolio level.
Every project may look plausible while all of them rest on the same mistaken
assumption.  Selection should reserve room for lines of inquiry that could
reveal that mistake.

A useful analogy comes from genetic algorithms.  A population can converge too
early when selection eliminates alternatives before they have been adequately
explored.  Scientific agents may face a similar problem when an initial framing
shapes every proposal that follows.  We hypothesize that using several models
on the same problem can preserve competing lines of inquiry, especially when
the models begin from different assumptions or methods.  URSA's agent symposia
provide one way to organize such exchanges
\citep{ursadocs2026symposia}.  Agreement among several models still cannot
establish that an idea is sound when the models share training data and failure
modes.

Facilities should reserve some capacity for experiments that could overturn
the assumptions guiding the rest of the program.  Otherwise an efficient
system may spend every available resource refining one mistaken view.

Scientific programs pursue several objectives at once.  Information gain may
favor a different experiment than urgency or mission consequence.  A rare,
high-consequence event may deserve attention even when its expected information
gain is modest.  These choices expose judgments that were easier to leave
implicit when expert time limited the number of proposals under consideration.

\section{The Physical Bottleneck: Reality Does Not Scale Like Computation}
\label{sec:physical}

Inference may get cheaper while accelerators still have finite beam time.
Laboratories can synthesize only so many materials.  Telescopes have limited
observing schedules.  Instruments, energy, manufacturing capacity, and compute
remain constrained.  Hypotheses may become abundant while access to reality
remains limited.

Physical access includes the work surrounding a measurement.  A specimen may
need preparation and a facility may require a safety review before an
instrument can run.  The resulting data are useful only if the instrument
state and experimental context were recorded.  Automation can reduce this
work, but it cannot make a unique material or observation available on demand.
Some missed observations cannot be repeated.

A validated simulation can rule out many candidates before they reach an
experiment, which raises the value of the remaining facility time.  Its
predictions depend on models, parameters, numerical approximations, and
validation data obtained from the world.  Both new measurements and
well-curated historical data become more valuable as simulation campaigns
expand, because they establish where modeled results can be trusted.

Closed-loop laboratories offer one response.  Agents can propose experiments,
control instruments within approved limits, analyze results, and choose a next
measurement.  The loop should include admission controls for safety and scarce
equipment, along with records of proposals that were rejected or failed.
Negative results will help later systems avoid repeating work, provided that
the conditions and uncertainty of those results have been preserved.

Closed-loop laboratories are most useful where specimens, protocols, and
instruments can be standardized.  They do less to relax physical constraints
in fields that depend on unique observations, field campaigns, or large
facilities.

\section{The Validation Bottleneck:
  From Producing Answers to Supporting Claims}
\label{sec:validation}

A model can produce a detailed answer in seconds.  The measurements needed to
support it may take weeks, and weak provenance can make an otherwise correct
result unusable.  Model inadequacy, correlated errors, contaminated literature,
and failures of reproducibility remain as reasoning gets cheaper.  Machine
reasoning may eventually extract far more from the available evidence than it
does today.  Verification and uncertainty characterization then determine how
far the evidence can support a claim.

A fluent explanation can combine many true statements and still make a final
claim that the evidence does not support.  Supporting a claim requires a
traceable chain from conclusion back to evidence.  A reviewer should be able
to find the relevant measurement and reproduce the analysis, including the
points where judgment entered.  Provenance is part of the result because it
shows why the conclusion deserves confidence.

Agents may produce claims faster than scientists can check them.  Only a small
fraction can receive human attention.  Scientific agents will therefore need
to recognize when a result deserves escalation and ask for review.  When a
system asks for review, it should begin by explaining why.  The scientist then
needs a concise account of how the agents reached that point and which evidence
remains unresolved.  This matters especially in multi-agent systems, where a
long internal debate can leave the supervisor unable to tell what the agents
concluded.  Before a result guides an experiment or mission decision, the
system must turn that history into a reviewable account.

Model capability is also uneven across tasks that look similar.  Dell'Acqua and
colleagues call this the \emph{jagged technological frontier}: AI assistance can
improve performance on one task and degrade it on another nearby task
\citep{dellacqua2026jagged}.  Each part of a scientific workflow therefore
needs its own evaluation.

Using several models can reveal disagreements and failure modes that one
self-reviewing model might conceal.  Shared training data and assumptions can
also lead several models to agree on the same error.  Claims produced through
agent debate still have to be tested against evidence.

Scientific systems will need calibrated records of where their methods work.
Performance on literature search says little about performance on uncertainty
quantification, and success in one physical regime may not transfer to another.
Evaluations should follow the actual workflow and include failures that occur
when several individually capable components interact.  Independent data,
alternative methods, and physical tests provide stronger checks than agreement
among agents built from similar foundations.

Human review is best spent on the assumptions, evidence, and consequences that
matter.  Outputs designed for inspection need explicit uncertainty and durable
links back to source material, so reviewers do not have to reconstruct an agent
run before they can evaluate its conclusion.

\section{The Institutional Bottleneck: Who Chooses What Science Gets Done?}
\label{sec:institutions}

Institutions inherit decisions that technical ranking cannot settle.  A
facility operates under priorities and risk tolerances that someone must
authorize.  That authority also carries responsibility for the evidence
required and for the outcome of an autonomous campaign.

An experiment that maximizes information gain may do little for an urgent
mission, and a high-value mission result may carry risks that the institution
will not accept.  Choosing among those aims requires institutional priorities.

Alvin Weinberg described scientific priority setting as a problem in the
\emph{axiology of science}: the value system used to decide which competent,
feasible science deserves support.  Automated selection does not remove that
problem.  It encodes judgments about scientific, technological, social, and
mission value in ranking rules and facility-admission policies
\citep{weinberg1963criteria,weinberg1970axiology}.

Once a priority is encoded in a ranking rule, it can shape an entire research
program.  Each rule therefore needs an institutional owner and a record of how
it changes when priorities conflict or new evidence arrives.

An autonomous system may receive operational authority without becoming an
accountable institutional actor.  Alignment and engineering controls can shape
its behavior, but they do not make the system's own future depend on its present
conduct \citep{debardeleben2026accountability}.  Institutions must therefore
decide how much authority can be exercised in the absence of accountability at
the point of action.  Bounded authority and independent review become part of
scientific system design.  An agent that drafts a proposal raises a different
governance question than one that reserves facility time or changes an
experimental control.  Clear boundaries should identify which actions are
advisory, which require approval, and which may be performed autonomously.
Consequential actions need a named institutional owner even when no person
selected each intermediate step.

Risk assessment also requires accountable authority.  Quantitative analysis
can characterize much of a risk without deciding whether the residual risk is
acceptable.  Consequential choices often involve uncertain or incomparable
harms, mission urgency, and institutional values.  Agents can prepare and
challenge the assessment.  An authorized person or body must accept
responsibility for the decision to proceed, pause, or stop.

Access will also become a scientific policy.  Agents may need controlled data,
specialized models, hazardous materials, or expensive instruments.  Broad
access can accelerate discovery, while unrestricted access can expose private
information or create unacceptable physical and security risks.  Institutions
will have to grant capabilities according to purpose and evidence of reliable
behavior, then retain records that support later review.

\section{The Changing Role of the National Laboratory}
\label{sec:labs}

If expert reasoning becomes widely available, facilities and controlled data
will account for more of what distinguishes a national laboratory.  Machine
systems can use those assets only through interfaces that preserve safety and
provenance.

The compute-only cases in Section~\ref{sec:compute-only} mark the limit of this
argument.  A proof or algorithm that never needs a facility can be produced
wherever models, compute, and expert review are available.  The laboratory's
advantage grows when scientific work must pass through distinctive data,
materials, instruments, or authorities.

\begin{claimbox}[title=Candidate Knowledge Is Not Trusted Knowledge]
AI may make candidate knowledge cheap without making trustworthy, actionable
knowledge cheap.  National laboratories help turn candidate results into
trusted evidence by combining machine reasoning with controlled experiments,
domain judgment, and accountable authority.
\end{claimbox}

This opportunity is conditional.  Laboratories need interfaces through which
machine systems can request resources and return evidence.  The institution
still sets admission criteria, experimental standards, authority limits, and
the conditions under which results enter the scientific record.  Release
reviews, access controls, and facility queues protect safety, security, and
public trust.  Their design will also determine whether agentic science can
use the laboratory's scarce assets at all.

This work reaches beyond operating facilities.  Laboratories can define how
agents describe experimental constraints and report results with uncertainty
and provenance.  Instruments and simulation codes will need machine-readable
descriptions of their capabilities.  The systems that expose them will need
controls strong enough for an agent to use them without bypassing safety,
security, or scientific review.

Laboratories can test models where it matters most: against physical
measurements.  A campaign that covers several operating regimes can establish
where predictions fail.  Preserving those failures gives later decisions
about autonomy an empirical basis.

Historical data may become newly valuable in this setting.  Records collected
for one program can test claims generated for another, especially when they
include calibration history and experimental context.  Curating those records
for machine use is scientific infrastructure work.  Weak metadata can make a
large archive less useful than a smaller collection whose provenance is clear.

The workforce must change with the technical system.  Routine synthesis,
coding, and analysis may require fewer person-hours.  Expertise will shift
toward experimental design, instrumentation, metrology, anomaly response,
evidence integration, safety and security, and accountable program leadership.
Scientists will still need to recognize when an agent has framed the wrong
problem, chosen an invalid model, or exceeded the evidence.  These roles
determine whether physical activity becomes trustworthy evidence.  Scientists
acquire that judgment partly through work that agents may now perform.
Laboratories should preserve forms of apprenticeship that give early-career
scientists direct experience with instruments, models, failed analyses, and
difficult decisions.

At Los Alamos National Laboratory, these roles are already taking shape.
ArtIMis is a large multidisciplinary effort to develop AI capabilities across
Laboratory missions, with explicit attention to data quality, uncertainty, and
trust in high-consequence settings \citep{lawrence2025artimis}.  The open source
\href{https://github.com/lanl/ursa}{URSA} system connects agents for planning,
research, hypothesis formation, code execution, domain tools, and advanced
physics simulations \citep{grosskopf2026ursa}.  Together, the projects show how
a national laboratory can adapt general AI capabilities to mission science,
then test and govern their use.

The quality of this connection will matter as much as the assets themselves.
General models will continue to improve, and many organizations will have
access to them.  Far fewer institutions can join those models to distinctive
facilities, trusted data, domain judgment, and authority while producing
evidence that others can audit.

\section{Designing Science for an Age of Abundant Intelligence}
\label{sec:design}

Consider a campaign to find a material that survives an extreme environment.
Agent teams generate ten thousand candidate compositions and use simulations
to reduce the set to forty.  A program chooses six for synthesis, and only two
fit the available instrument schedule.  The measurements then have to be tied
to calibration records, processing history, and model assumptions before a
composition can be declared successful.  A project leader may still have to
choose between the candidate with greatest information value and the one most
relevant to an urgent mission.  Every handoff spends a different scarce
resource.

The campaign needs an architecture built around those handoffs.  Simulation
and instrument interfaces expose capabilities, limits, and cost in forms an
agent can use.  Each experiment request should explain why the measurement
deserves facility time.  It should identify the uncertainty at stake and say
how either result would change the campaign.  A scheduler can then compare
requests using the program's actual objectives.

Running an experiment should not automatically place its result in the
scientific record.  Agents may execute low-risk work on their own, while a
separate review decides what the program accepts as evidence.  This boundary
can move as evidence about the system accumulates.

Provenance begins with the first agent action.  Reconstructing it after a long
run is unreliable, particularly when the run includes changing code, retrieved
sources, simulation inputs, and intermediate judgments.  The record connects
each consequential claim to the data and operations that support it, along with
uncertainty and known limitations in forms that later machines can inspect.

Evaluation follows the complete path from proposal to measurement to
conclusion.  A model benchmark cannot establish that an autonomous experimental
campaign is safe or scientifically sound.  Tests need conditions that invite a
plausible error and reveal whether the system recognizes inadequate evidence.

Review queues, publication incentives, and funding processes can be overwhelmed
by material that is cheap to generate.  Institutions can reward evidence,
reproducibility, and useful reductions in uncertainty while training scientists
to supervise automated work and examine its physical and mathematical
foundations.

Human supervision becomes difficult when agents work much faster than their
reviewers.  A person may check the work periodically, but an agent team can
make thousands of decisions between those checks.  Reviewing every step would
erase much of the advantage.  The system must instead identify the decisions
that need human judgment.

Mitchell and colleagues argue that the fast, dense stream of agent actions can
overload overseers and reduce review to routine approval
\citep{mitchell2026agents}.  Controls outside the agent must be able to pause
work at machine speed.  They must also preserve enough context for a reviewer
to understand what happened.  Supervision matters only if a person sees the
decision in time to redirect the work.

The transition can begin before the premise of this paper is fully realized.
Well-bounded interfaces, complete provenance, and explicit selection criteria
are useful under present conditions.  As the volume of machine generated work
grows, they will determine whether abundant reasoning produces reliable
science or simply overwhelms the next queue.

\section{Conclusion: After the Cognitive Bottleneck}
\label{sec:conclusion}

The scarcity inversion will arrive unevenly as machine reasoning 
is already becoming plentiful in
software and mathematics while remaining limited in fields that depend on
specialized observations or specialized knowledge.  
However, even a moderate change could
create a severe mismatch when one part of the workflow generates work faster
than the next part can process it.

Compute-only fields are moving first - the progression from program search to
new mathematical constructions, formal proofs, and the recent Navier--Stokes
claim suggests that some research areas may change before empirical science
has comparable systems.  The remaining challenges are compute allocation,
verification, expert review, and institutional acceptance.

The practical test for an organization is which part of its workflow is already
producing work faster than the next part can absorb, and which resource now
sets the pace.  The answer may differ across programs inside the same
laboratory.  We are already seeing this with swarms of AI agents that produce
information at a rate beyond what humans can interpret.

If this argument is right, model throughput is a poor measure of scientific
capacity.  A laboratory could field millions of agent-hours and make little
progress because its experiments, evidence, review, or authority cannot absorb
the output.  Useful measures should instead ask whether a program chooses
informative experiments and converts their results into validated conclusions.

National laboratories have an unusual opportunity because they combine
advanced computing with facilities, specialized data, controlled materials,
and public authority.  Compute-only work remains part of their mission and may
draw on information or expertise that cannot be assembled elsewhere.  The
laboratory becomes especially valuable when machine reasoning has to be tested
against measurements or used in decisions for which the institution is
accountable.  This requires usable interfaces to facilities and data, clear
authority, and evidence that others can audit.  The laboratory owns the part of
the scientific loop that faces the physical world.  It decides which
machine-generated proposals deserve scarce resources, makes sure measurements
can be interpreted, determines what claims the evidence supports, and accepts
responsibility for consequential action.  Abundant reasoning can expand the
questions a laboratory pursues.  Progress still depends on choosing which
questions to test against evidence.

\bibliographystyle{plainnat}
\bibliography{references}

\end{document}